\documentclass[journal]{article}
\usepackage[margin=1 in]{geometry}

\usepackage[utf8]{inputenc}
\usepackage[T1]{fontenc}
\usepackage{etex}
\usepackage{graphicx}
\usepackage{fleqn}
\usepackage{cite}
\usepackage{physics}
\usepackage{siunitx}
\usepackage{amsmath,amssymb,amsfonts}
\usepackage{algorithmic}
\usepackage{graphics}
\usepackage{textcomp}
\usepackage{tikz}
\usepackage{relsize}
\usepackage{color}
\usetikzlibrary{arrows, arrows.meta}
\usepackage{comment}
\usepackage{xcolor}
\usepackage{filecontents}
\usepackage{float}
\usetikzlibrary{arrows.meta,patterns}
\usetikzlibrary{shapes.misc}
\usepackage{graphics}
\usepackage{pict2e,color}
\usepackage{bbm}
\usepackage{subcaption}
\usepackage{nccmath}
\usetikzlibrary{fit}
\usepackage{nccmath}
\usepackage{pgfplots}
\usepackage{lipsum}
\usepackage{stackengine}
\usepackage[capitalize,noabbrev]{cleveref}
\usepackage[ruled,vlined]{algorithm2e}
\usepackage{amsthm}
\usepackage{balance}
\usepackage{color,soul}
\usepackage{dsfont}

\setul{0.5ex}{0.3ex}
\setulcolor{blue}
\def\BibTeX{{\rm B\kern-.05em{\sc i\kern-.025em b}\kern-.08em
    T\kern-.1667em\lower.7ex\hbox{E}\kern-.125emX}}

\def\E{\mathrm{E}}
\def\lim{\mathrm{lim}}

\def\dB{\mathrm{dB}}

\def\e{\mathrm{e}}

\def\ln{\mathrm{ln}}

\usepackage{acro}
\DeclareAcronym{NN}{short=NN, long=neural network}
\DeclareAcronym{FFNN}{short=FFNN, long=feed-forward neural network}
\DeclareAcronym{GNN}{short=GNN, long=graph neural network}
\DeclareAcronym{GGNN}{short=GGNN, long=gated graph neural network}
\DeclareAcronym{GRU}{short=GRU, long=gated recurrent unit}
\DeclareAcronym{RL}{short=RL, long=reinforcement learning}

\usepackage[colorinlistoftodos,prependcaption,textsize=footnotesize]{todonotes}
\usepackage{regexpatch}
\makeatletter
\xpatchcmd{\@todo}{\setkeys{todonotes}{#1}}{\setkeys{todonotes}{inline,#1}}{}{}
\makeatother

\begin{document}

\title{On the Fidelity Distribution \\ of Link-level Entanglements under Purification}
 \author{Karim Elsayed$^*$, Wasiur R. KhudaBukhsh, Amr Rizk$^*$  \\
\textit{$^*$Faculty of Computer Science, University of Duisburg-Essen,
Germany }\\
\textit{School of Mathematical Sciences, University of Nottingham, UK} 
}
\date{}
\maketitle

\begin{abstract}
Quantum entanglement is the key to quantum communications over considerable distances.
The first step for entanglement distribution among quantum communication nodes is to generate link-level Einstein–Podolsky–Rosen (EPR) pairs between adjacent communication nodes.
EPR pairs may be continuously generated and stored in a few quantum memories to be ready for utilization by quantum applications.
A major challenge is that qubits suffer from unavoidable noise due to their interaction with the environment, which is called decoherence.
This decoherence results in the known exponential decay model of the fidelity of the qubits with time, thus, limiting the lifetime of a qubit in a quantum memory and the performance of quantum applications.

In this paper, we evaluate the fidelity of the stored EPR pairs under two opposite dynamical and probabilistic phenomena, first, the aforementioned decoherence and second purification, i.e. an operation to improve the fidelity of an EPR pair at the expense of sacrificing another EPR pair.
Instead of applying the purification as soon as two EPR pairs are generated, we introduce a Purification scheme Beyond the Generation time (PBG)  of two EPR pairs.
We analytically show the probability distribution of the fidelity of stored link-level EPR pairs in a system with two quantum memories at each node allowing a maximum of two stored EPR pairs.
In addition, we apply a PBG scheme that purifies the two stored EPR pairs upon the generation of an additional one.
We finally provide numerical evaluations of the analytical approach and show the fidelity-rate trade-off of the considered purification scheme.

\end{abstract}
\section{Introduction}
Quantum entanglement lies at the core of the quantum Internet which enables quantum applications including quantum communications~\cite{cacciapuoti_Caleffi_entang_meets_calssical,deng_qu_comm}, quantum key distribution~\cite{Ursin_longdist_keydist,epping_qu_key_dist} and distributed quantum computation~\cite{Caleffi_QInternet}.
A major challenge is that qubits suffer from unavoidable decoherence,  which results in a rapid decay in the quality of the entangled Einstein–Podolsky–Rosen (EPR) qubit pair with time~\cite{cacciapuoti_qu_decoherence,schlosshauer_qu_deocherence}.
A corresponding quality metric, also denoted fidelity, measures the closeness between the noisy EPR pairs and the original (desired) one.
In the phase damping decoherence model, the fidelity decays exponentially with time~\cite{nielsen_QC_QI_book}.

\begin{figure}[t]
\centering
     {\input{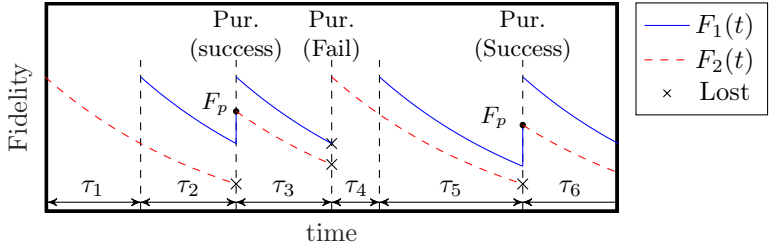}}    
    \caption{A sample path realization of the fidelity of two EPR pairs stored in the size-two memory system with fidelities $F_1(t)$ and $F_2(t)$ such that $F_2(t)$ always represents the lower fidelity EPR pair. Entanglement inter-generation times are given by the random sequence $\{\tau_i\}$. Purification takes place upon a new entanglement generation \textit{to a full system} where the fidelity improves in case of purification success while the two stored pairs are lost in case of purification failure.}
        \label{fig:fidelity_sketch}
\end{figure}

A canonical model for quantum networks with quantum memories or queues assumes that EPR pairs are continuously generated and stored to be ready to respond to transmission requests of qubits resulting in a high-capacity network \cite{dahlberg_qu_Linllayer_protocol}.
The quantum network must guarantee a sufficient fidelity for the desired application and due to the probabilistic nature of quantum operations the higher the fidelity, the better the quality attained by the application.
To this end, the goal of network nodes is to generate high-fidelity entanglements, ensure the validity of the stored ones and apply purification to them. 

The generation of high-fidelity entanglements through purification involves consuming a smaller or equal fidelity EPR pair to improve the fidelity of another pair.
In~\cite{Inside_quantum_repeater,Ruan_pur_protocol,pan_entanglement_pur,deutsch_quantum_pur} different recurrence purification schemes are proposed using multiple purification rounds to generate one very high-fidelity EPR pair.
Specifically, we start from the purification scheme in~\cite{Ruan_pur_protocol} as a baseline to compute the fidelity distribution. 
Also note that some works consider entanglement cut-off times, i.e., a deadline after which the fidelity is assumed below a required threshold, to ensure a minimum validity of the stored EPR pairs. 
For example, the work in \cite{Towsley_stochastic_qu_swithc} assumes the cut-off times are probabilistic and modeled by an exponential distribution based on which the EPR pairs stored in the quantum queue are dropped.

In this paper, we address the gap in the literature on the derivation of the fidelity steady-state distribution of stored EPR pairs under purification.
Purification is usually treated as a mechanism to initially generate high fidelity EPR pairs~\cite{Inside_quantum_repeater,Ruan_pur_protocol,pan_entanglement_pur,deutsch_quantum_pur}. 
Its application beyond the initial generation on the stored EPR pairs is rarely considered.
The purification of the stored EPR pairs has the potential to improve the fidelity at the expense of reducing the average number of EPR pairs in the system, which inherently leads to a rate-fidelity trade-off.
We denote the purification scheme applied beyond the generation of an EPR pair as PBG.

In this paper, we derive the steady-state probability distribution of the fidelity of the link-level entanglements in a system with a few quantum memories in isolation of any request process.
In addition, we apply a PBG scheme that distillates the stored EPR pairs before storing a newly generated pair when the quantum memory is full.
To the best of our knowledge, this is the first work that evaluates the fidelity distribution of the stored EPR pairs in the quantum memories and the effect of the PBG schemes.

The remainder of the paper is structured as follows: We first describe the model and problem statement in Sect.~\ref{sect:model}. 
In Sect.~\ref{sect:approach}, we derive the steady-state fidelity distribution of the stored EPR pairs. 
We numerically evaluate the proposed approach in Sect.~\ref{sect:numerical_validation} and summarize the related work in Sect.~\ref{sect:related_work} before concluding the paper and discussing open problems in Sect.~\ref{sect:Discussion}.

\section{Model and Problem statement}
\label{sect:model}




We model the entanglement generation as Bernoulli trials with success probability $p_g$ within a time slot $\triangle t$ similar to~\cite{Twosley_ideal_Qu_switch,Dai_Qu_Queuing_delay}.
One rationale that the entanglement generation is probabilistic is that the optical fiber is assumed to absorb the transmitted qubit from one node to the other with probability $1-p_g=1-\e^{-\eta l}$, where $l$ is the fiber length between the communication nodes and $\eta$ is the attenuation coefficient~\cite{guha_rateloss_Qu_repeater}.
This is associated with the link-level entanglement generation schemes that require qubit transmission through a fiber of length $l$ as discussed in~\cite{Inside_quantum_repeater,cacciapuoti_Caleffi_entang_meets_calssical}.
The scheme that we consider in this paper involves, first, the preparation of an EPR pair at one node before sending half of it, i.e., one of the two entangled qubits, to the other node, hence, $l$ denotes the link length.
In addition, each entanglement generation attempt takes ideally $\triangle t= l/c$ duration, where $c$ is the speed of light.
Following the formulation from~\cite{Inside_quantum_repeater} the fidelity of an EPR pair at time $t_0$ decays with time due to decoherence as
\begin{equation}
    F(t)= \frac{1}{2}\left(1+\left(2F(t_0)-1\right) \e^{-(t-t_0)/t_c}\right) ,
    \label{eq:fidleity_ent_cont}
\end{equation}
where $1/t_c$ is the decoherence rate and $F(t_0)$ is the fidelity of the EPR pair at time $t_0$.
In this work, we assume a perfect EPR generation.
Note that this assumption does not affect our analytical approach to obtain the fidelity distribution. 

We assume a PBG scheme to maintain high fidelity.
This entails attempting to purify the two stored EPR pairs at the moment of a successful generation of an additional one. 
Instead of dropping the lowest fidelity EPR pair to be replaced by the freshly generated one, we use it to purify the other stored EPR pair.
Specifically, we consider the purification scheme in~\cite{Ruan_pur_protocol}, where the fidelity of the purified EPR pair becomes
\begin{equation}
    F_p(F_1,F_2)=\frac{F_1 F_2}{F_1 F_2+(1-F_1)(1-F_2)},
    \label{eq:pur_fidelity}
\end{equation}
with a purification success probability $p_s$ given by
\begin{equation}
    p_s(F_1,F_2)=F_1 F_2+(1-F_1)(1-F_2).
    \label{eq:pur_success}
\end{equation}
Here, $F_1$ and $ F_2$ denote the fidelity of the first and the second pair, respectively. 

In Fig.~\ref{fig:fidelity_sketch}, we illustrate the model of the purification protocol using a sample path realization of the fidelity of the EPR pairs over time.
We assume the system contains one EPR pair at time $t=0$ and its fidelity decays with time due to decoherence as in \eqref{eq:fidleity_ent_cont}.
As per the Bernoulli assumption on the generation from above the inter-generation times $\{\tau_i\}_i$ come from a geometric distribution denoting the time between two successful EPR pair generations. 
When an EPR pair is generated and the quantum memories are full, purification takes place between the two stored EPR pairs.
The figure shows the improved fidelity obtained from purification as well as the random event of purification failure leading to losing the two stored EPR pairs.

Next, we calculate the steady-state distribution of the fidelity of the EPR pairs in the system with a few quantum memories.
The hardness of the problem originates from the hardness of tracking the fidelity due to its dependence on the purification outcome which in turn recursively depends on the fidelity at the previous purification attempts.

\section{Approach}
\label{sect:approach}
\begin{figure}[t]
\centering
     {\input{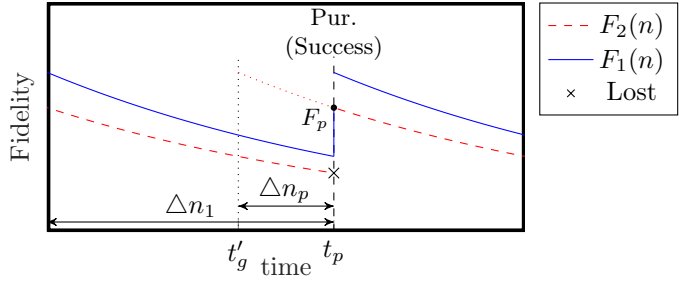}}    
    \caption{A sample path realization showing the application of PBG on the two stored EPR pairs with ages $\triangle n_1$ (shown) and $\triangle n_2$ (not shown), respectively, obtains a purified EPR pair with a fidelity value equivalent to an EPR generated at a later time point with shortened age $\triangle n_p < \triangle n_1$. The time point $t_g'$ is the hypothetical generation time of the EPR pair with an equivalent fidelity to the purified EPR pair at $t_p$.}
        \label{fig:virtual_age}
\end{figure}
Motivated by the Bernoulli modeling of the EPR generation in Sect.~\ref{sect:model}, our key idea for calculating the fidelity distribution is to track the fidelity decay at each time slot by discretizing the fidelity proportional to the time slots. 
This allows modeling the fidelity using a discrete time Markov chain (DTMC). 
We divide the fidelity range into $N+1$ discrete levels  proportional to its decay ranging from the lowest fidelity value $F_\epsilon$ to the initial fidelity of the generated EPR pair $F_0$ as
\begin{equation}
    F(\triangle n)=\frac{1}{2} \left(1+(2F_0-1) \e^{-\alpha \triangle n }\right),
    \label{eq:discrete_fidelity}
\end{equation}
where $ \triangle n \in \{0,1,...,N\}$ \textit{is the time duration elapsed since the  entanglement generation}, which we denote \textit{the age} (given in discrete time) and $\alpha:=\triangle t / t_c$ denotes the decoherence coefficient in one time slot.
We do not consider the fidelity beyond the lowest value $F_\epsilon$.
Since the age $\triangle n$ uniquely defines the fidelity level, we model the fidelity level as a result of a successful purification of two EPR pairs by an EPR pair with equal or smaller age $ \triangle n_p$ according to
\begin{align}
     & \triangle n_p(\triangle n_1,\triangle n_2) = \nonumber \\
    &\max \left(\left\lceil{ \frac{-1}{\alpha}\ln \left(\frac{2F_p(\triangle n_1,\triangle n_2)-1}{2F_0-1}\right)}  \right\rceil ,0 \right) ,
    \label{eq:pur_quantized_age}
\end{align}
where $F_p(\triangle n_1,\triangle n_2)$ is the fidelity after purification of the two EPR pairs from \eqref{eq:pur_fidelity} and $\triangle n_1$ and $\triangle n_2$ are the ages corresponding to the fidelities of the stored EPR pairs $F_1(n)$ and $F_2(n)$, respectively.
Here $F_i(n)$ is the fidelity at slot $n$ on the discrete time lattice. 
Since $F_p(\triangle n_1,\triangle n_2)$ may not correspond to one of the discrete fidelity levels, we use $\lceil . \rceil$ to map the \textit{purification age} to the next larger integer to lower bound the purified fidelity.
In case $F_p>F_0$, which may occur for small initial EPR fidelity, the maximum operation in \eqref{eq:pur_quantized_age} maintains $\triangle n_p\geq 0$ corresponding to the highest fidelity~$F_0$. 

Note that the reduced age due to purification does not reflect the actual time the EPR pair spent in the memory.
In our model, the purified pair obtains a fidelity value from \eqref{eq:pur_fidelity} with success probability \eqref{eq:pur_success} that is equivalent to the fidelity of an EPR with a later generation time. 
Hence, as shown in \ref{fig:virtual_age} the age $\triangle n$ is shortened accordingly through the purification operation. 

Similarly, we calculate the maximum age $N$ that achieves the lowest fidelity threshold according to $F(N)=F_\epsilon$ as
\begin{equation}
    N= \left\lceil \frac{-1}{\alpha} \ln\left(\frac{2F_\epsilon-1}{2F_0-1} \right)\right\rceil.
\end{equation}

Note that the fidelity $F_1(n)$ always represents the larger fidelity EPR pair out of the two stored ones when the memories are full and is exactly calculated using \eqref{eq:discrete_fidelity}.
Hence, right after a fidelity jump in Fig.~\ref{fig:fidelity_sketch}, $F_2(n)$ represents the older EPR pair and the fidelity of the only EPR pair in the system when it is not full (cf. the figure).
The value of $F_2(n)$ is quantized according to \eqref{eq:pur_quantized_age} during purification.
\subsection{DTMC model of the age of the stored EPR pairs}

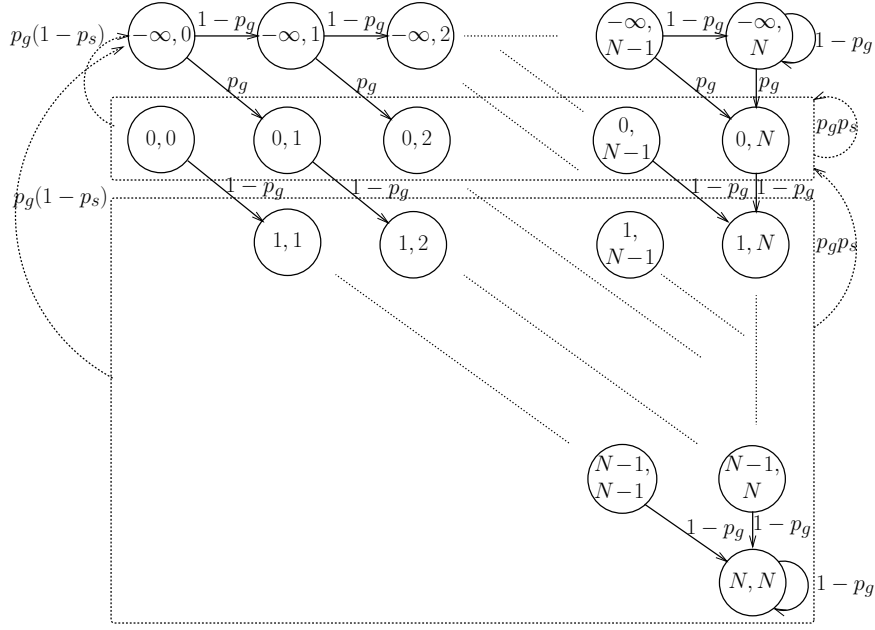
\begin{figure}[t]
\centering
     \resizebox{0.7\linewidth}{!}{{\pgfkeys{/pgf/fpu/.try=false}%
\ifx\XFigwidth\undefined\dimen1=0pt\else\dimen1\XFigwidth\fi
\divide\dimen1 by 11056
\ifx\XFigheight\undefined\dimen3=0pt\else\dimen3\XFigheight\fi
\divide\dimen3 by 7771
\ifdim\dimen1=0pt\ifdim\dimen3=0pt\dimen1=4143sp\dimen3\dimen1
  \else\dimen1\dimen3\fi\else\ifdim\dimen3=0pt\dimen3\dimen1\fi\fi
}
\begin{tikzpicture}[x=0.16, y=0.16, every node/.style={scale=1.8}]
{\ifx\XFigu\undefined\catcode`\@11
\def\temp{\alloc@1\dimen\dimendef\insc@unt}\temp\XFigu\catcode`\@12\fi}
\XFigu4143sp
\ifdim\XFigu<0pt\XFigu-\XFigu\fi
\pgfdeclarearrow{
  name = xfiga0,
  parameters = {
    \the\pgfarrowlinewidth \the\pgfarrowlength \the\pgfarrowwidth},
  defaults = {
	  line width=+7.5\XFigu, length=+360\XFigu, width=+200\XFigu},
  setup code = {
    \dimen7 2.15\pgfarrowlength\pgfmathveclen{\the\dimen7}{\the\pgfarrowwidth}
    \dimen7 2\pgfarrowwidth\pgfmathdivide{\pgfmathresult}{\the\dimen7}
    \dimen7 \pgfmathresult\pgfarrowlinewidth
    \pgfarrowssettipend{+\dimen7}
    \pgfarrowssetbackend{+-\pgfarrowlength}
    \dimen9 -0.5\pgfarrowlinewidth
    \pgfarrowssetvisualbackend{+\dimen9}
    \pgfarrowssetlineend{+-0.5\pgfarrowlinewidth}
    \pgfarrowshullpoint{+\dimen7}{+0pt}
    \pgfarrowsupperhullpoint{+-\pgfarrowlength}{+0.5\pgfarrowwidth}
    \pgfarrowssavethe\pgfarrowlinewidth
    \pgfarrowssavethe\pgfarrowlength
    \pgfarrowssavethe\pgfarrowwidth
  },
  drawing code = {\pgfsetdash{}{+0pt}
    \ifdim\pgfarrowlinewidth=\pgflinewidth\else\pgfsetlinewidth{+\pgfarrowlinewidth}\fi
    \pgfpathmoveto{\pgfqpoint{-\pgfarrowlength}{0.5\pgfarrowwidth}}
    \pgfpathlineto{\pgfqpoint{0pt}{0pt}}
    \pgfpathlineto{\pgfqpoint{-\pgfarrowlength}{-0.5\pgfarrowwidth}}
    \pgfusepathqstroke
  }
}

\clip(-952,-11667) rectangle (9880,-3896);
\tikzset{inner sep=+0pt, outer sep=+0pt}
\pgfsetlinewidth{+40\XFigu}
\pgfsetstrokecolor{black}
\pgfsetdash{}{+0pt}
\pgfsetarrows{[line width=7.5\XFigu]}
\pgfsetarrowsend{xfiga0}
\draw (7020,-10170)--node[right,yshift=3pt, font=\Huge]{$1-p_g$}(8055,-10890);
\draw (8370,-10260)--node[right,yshift=3pt, font=\Huge]{$1-p_g$}(8370,-10710);
\pgfsetfillcolor{black}
\pgfsetdash{}{+0pt}
\draw (8685,-10890) node[below,yshift=-15pt,xshift=75pt, font=\Huge]{$1-p_g$} arc[start angle=+103.3, end angle=+-112.1, radius=+307.9];
\draw (8775,-4050)  node[below,yshift=-20pt,xshift=65pt, font=\Huge]{$1-p_g$} arc[start angle=+103.3, end angle=+-112.1, radius=+307.9];
\pgfsetdash{{+60\XFigu}{+60\XFigu}}{++0pt}
\draw (9135,-7965)  node[right, yshift=90pt, xshift=3pt, font=\Huge]{$p_gp_s$} arc[start angle=+-56.05, end angle=+56.05, radius=+1200.2];
\draw (9135,-5805)  node[right, yshift=30pt, xshift=3pt, font=\Huge]{$p_gp_s$}arc[start angle=+-114.8, end angle=+114.8, radius=+380.2];
\draw (400,-5445)  node[left, yshift=100pt, xshift=-10pt, font=\Huge]{$p_g(1-p_s)$}arc[start angle=+249.19, end angle=+92.19, radius=+581.6];
\draw (380,-8595) node[left,yshift=200pt, xshift=-5pt, font=\Huge]{$p_g(1-p_s)$} arc[start angle=+241.58, end angle=+114.68, radius=+2315.3];
\pgfsetdash{}{+0pt}
\draw  (6743,-9854) circle  [radius=+430,align=left] node [font=\Huge] {\begin{tabular}{l}
    $N\!-\!1,$ \\
    $N\!-\!1$
\end{tabular}}; 
\draw  (8363,-9854) circle [radius=+415] node [font=\Huge] {\begin{tabular}{c}
    $N\!-\!1,$ \\
    $N$
\end{tabular}}; 
\draw  (8370,-11150) circle [radius=+415] node [font=\Huge] {$N, N$};
\draw  (990,-4320) circle [radius=+415] node [font=\Huge] {$-\infty, 0$};
\draw  (2558,-5624) circle [radius=+415] node [font=\Huge] {$0, 1$};
\draw  (4178,-5624) circle [radius=+415] node [font=\Huge] {$0, 2$};
\draw  (4223,-4319) circle [radius=+415] node [font=\Huge] [font=\Huge] {$-\infty, 2$};
\draw  (2603,-4319) circle [radius=+415] node [font=\Huge] {$-\infty, 1$};
\draw  (984,-5614) circle [radius=+415] node [font=\Huge] {$0, 0$};
\draw  (8408,-5624) circle [radius=+415] node [font=\Huge] {$0, N$};
\draw  (6833,-4319) circle [radius=+415]node [font=\Huge] {\begin{tabular}{c}
    $-\infty,$ \\
    $N\!-\!1$
\end{tabular}}; 
\draw  (8453,-4319) circle [radius=+415]node [font=\Huge] {\begin{tabular}{c}
    $-\infty,$ \\
    $N$
\end{tabular}};
\draw  (8408,-6929) circle [radius=+415]node [font=\Huge] {$1, N$};
\draw  (4133,-6929) circle [radius=+415]node [font=\Huge] {$1, 2$};
\draw  (2563,-6898) circle [radius=+415]node [font=\Huge] {$1, 1$};
\draw  (6795,-5625) circle [radius=+415]node [font=\Huge] {\begin{tabular}{c}
    $0,$ \\
    $N\!-\!1$
\end{tabular}}; 
\draw  (6840,-6930) circle [radius=+415]node [font=\Huge] {\begin{tabular}{c}
    $1,$ \\
    $N\!-\!1$
\end{tabular}}; 
\draw (1395,-4320)-- node[above, yshift=3pt, font=\Huge]{$1-p_g$}(2205,-4320);
\draw (3015,-4320)--node[above,yshift=3pt, font=\Huge]{$1-p_g$} (3825,-4320 );
\draw (2880,-5895)--node[right,yshift=3pt, font=\Huge]{$1-p_g$}(3825,-6660);
\draw (2916,-4602)--node[right,yshift=3pt,xshift=3pt, font=\Huge]{$p_g$}(3861,-5367);
\draw (1318,-5879)--node[right,yshift=3pt, font=\Huge]{$1-p_g$}(2263,-6644);
\pgfsetlinewidth{+40\XFigu}
\pgfsetdash{{+40\XFigu}{+180\XFigu}}{+15\XFigu}
\pgfsetarrowsend{}
\pgfsetlinewidth{+40\XFigu}
\pgfsetdash{}{+0pt}
\pgfsetarrowsend{xfiga0}
\draw (7245,-4320)--node[above,yshift=3pt, font=\Huge]{$1-p_g$}(8055,-4320);
\draw (7154,-4592)--node[right,yshift=3pt,xshift=3pt, font=\Huge]{$p_g$}(8099,-5357);
\draw (7122,-5881)--node[right,yshift=3pt, font=\Huge]{$1-p_g$}(8067,-6646);
\pgfsetdash{}{+0pt}
\draw (8415,-4725)--node[right,yshift=3pt,xshift=3pt, font=\Huge]{$p_g$} (8415,-5220);
\draw (8415,-6030)--node[right,yshift=3pt, font=\Huge]{$1-p_g$}(8415,-6525);
\pgfsetlinewidth{+40\XFigu}
\pgfsetdash{{+40\XFigu}{+68\XFigu}}{+15\XFigu}
\pgfsetarrowsend{}
\draw (3163,-7300)--(6088,-9415);
\draw (4765,-7308)--(7690,-9423);
\draw (4810,-6228)--(7735,-8343);
\draw (4752,-4914)--(6192,-5994);
\draw (5220,-4500)--(6255,-5265);
\draw (8415,-7560)--(8415,-9180);
\draw (7228,-7297)--(8263,-8062);
\draw (4770,-4320)--(5940,-4320);
\pgfsetlinewidth{+40\XFigu}
\pgfsetdash{}{+0pt}
\pgfsetarrowsend{xfiga0}
\draw (1305,-4590)--node[right,yshift=3pt,xshift=3pt, font=\Huge]{$p_g$}(2250,-5355);
\pgfsetdash{{+60\XFigu}{+60\XFigu}}{++0pt}
\draw (9135,-11655) [rounded corners=+105\XFigu] rectangle (360,-6345);
 \draw (9135,-5085) [rounded corners=+105\XFigu] rectangle (360,-6120);
\end{tikzpicture}%
}  
    \caption{DTMC modelling of the fidelity of the EPR pairs stored in the system. A state $(x,y)$ corresponds to the age of the stored EPR pairs $x,y$, respectively where $x\leq y$ denotes the ordering of the pairs. Recall that the discretization in \eqref{eq:discrete_fidelity} provides a one to one mapping of the age to the discrete fidelity levels. The description of the transitions represented by the dotted arrows is in the text below.}
 \label{fig:DTMC}
\end{figure}

We model the fidelity of the EPR pairs stored in the system by a DTMC with states $(\triangle n_1,\triangle n_2) \sim (F_1(n),F_2(n))$ representing their age such that $\triangle n_2$ always represents the oldest (smallest fidelity) EPR pair in the system.
We assume that the system has initially one EPR pair with perfect fidelity, thus the initial system state is $(-\infty,0)$ at time $n=0$, where $-\infty$ stands for the non-existing second EPR pair.
We illustrate the system DTMC in Fig.~\ref{fig:DTMC}, where we denote the state transitions to be either \textit{forward} or \textit{backward}.
The forward transitions represent the time evolution before attempting purification, i.e., the age progression of EPR pairs. 
We summarize the forward transitions as
\begin{align}
    (i,j) &\rightarrow \left(\mathrm{min}\{i+1,N\},\mathrm{min} \left\{j+1,N \right\}\right) \text{w.p. } 1-p_g ,
     \nonumber \\
     & (-\infty,j ) \rightarrow  \left(0,\mathrm{min}  \left\{j+1,N \right\} \right) \text{w.p. } p_g .
     \label{eq:forward_transitions}
\end{align}
The backward transitions are a result of a purification attempt as in 
\begin{align}
    \text{Success: }(i,j ) \rightarrow &\left(0,\triangle n_p(i,j) \right) \text{w.p. }  p_g \ p_s(i,j), \nonumber \\
    \text{Fail: } (i,j) \rightarrow &\left(-\infty,0 \right) \text{w.p. }  p_g (1-p_s(i,j)), \nonumber \\
     & \forall i \in \left\{0,1, ..., N\right\}, \ j\geq i ,
     \label{eq:backward_transitions}
\end{align}
where $\triangle n_p(i,j)$ and $ p_s(i,j)$ are the age of the successfully purified EPR pair~\eqref{eq:pur_quantized_age} and the probability of purification success~\eqref{eq:pur_success} at state $(i,j)$, respectively.
The purification attempt occurs upon entanglement generation subject to a full quantum memory, thus it only appears when $i\neq -\infty$.
In case of a purification failure, the two stored EPR pairs are lost and only the newly generated EPR pair remains, thus the system state resets to $(-\infty,0)$.

The backward transition probabilities are state-dependent since the success probability depends on the fidelity levels \eqref{eq:pur_success}, i.e., the age and $\alpha$.
We describe this dependence in Fig.~\ref{fig:DTMC} by the dotted arrow representing the existence of a state-dependent transition from each state within a block to a corresponding state in the destination block. 
We define a \textit{block} in Fig.~\ref{fig:DTMC} to comprise the states within a horizontal row which represents the states $S_m=\{(m,j)\} \ \forall j\geq m, j \in \{0,1,..,N\}$.
Note that not only do the transition probabilities vary in the case of successful purification but also the destination state $(0,\triangle n_p(i,j))$.
As illustrated the destination state is a function of the current state as well as $\alpha$.
Note that a careful choice of $\alpha$, i.e., the time discretization with respect to the decoherence rate is crucial for the design of the DTMC.

We represent the transition matrix of this Markov chain in terms of sub-matrices describing the transitions between blocks of the DTMC as depicted in Fig.~\ref{fig:DTMC} with the states ordered as 
\\$[(-\infty,0) \dots (-\infty,N) (0,0) \dots (0,N) \dots \dots (N,N) ]$ by
\begin{equation}
\mathbf{Q}=
\small
    \begin{bmatrix}
        \pmb{0}_{N+1,1} \mathbf{X}_0 &\mathbf{X}_{-\infty} & \pmb{0}_{N+1,N} &\dots & \dots& 0
        \\
        \pmb{f}_0 \ \pmb{0}_{N+1,N}& \mathbf{D}_0 & \mathbf{X}_0 & \ddots &\ddots & \vdots
        \\
        \pmb{f}_1 \  \pmb{0}_{N,N}&\mathbf{D}_1 &  \pmb{0}_{N,N} & \mathbf{X}_1& \ddots &\vdots
        \\
        \vdots  & \vdots & \vdots & \ddots & \ddots& 0 
        \\
        \vdots &\vdots& \vdots&\ddots&\pmb{0}_{2,2} & \mathbf{X}_{N-1} 
        \\
        \pmb{f}_N \ \pmb{0}_{1,N} &\mathbf{D}_N & \pmb{0}_{1,N}&\dots&\pmb{0}_{1,2} & 1-p_g 
    \end{bmatrix}
    \label{eq:transition_matrix}
    .
\end{equation}
The forward transition implies the transition from one block to the next one, thus resulting in the sparse matrix structure, where 
$\mathbf{X}_m$ is an $N-m+1 \times N-m$ matrix, with $0 \leq m \leq N-1$, representing the forward transitions in \eqref{eq:forward_transitions}.
We express this matrix as 
\begin{equation}
\mathbf{X}_m= (1-p_g)
    \begin{bmatrix}
         \hspace{0.5 cm}\mathbf{I}_{N-m} \\
        \pmb{0}_{1,N-m-1} & 1
    \end{bmatrix}
    , 0 \leq m \leq N-1 ,
\end{equation}
while the $N+1 \times N+1$ matrix $\mathbf{X}_{-\infty}$ represents the forward transitions due to the successful generation of an EPR pair when only one EPR pair is stored, which we express as
\begin{equation}
     \mathbf{X}_{-\infty}=\left[\pmb{0}_{N+1,1}| 1-\mathbf{X}_0\right] ,
\end{equation}
where $[.|.]$ represents the column-wise concatenation operation.
The probabilities of entanglement generation resulting in a failed purification attempt, thus the backward transitions in~\eqref{eq:backward_transitions}, are represented by the $N-m+1 \times 1$ vectors 
\begin{equation}  
\pmb{f}_m:=p_g\left(\pmb{1}-\left[p_s(m,m),p_s(m,m+1),...., p_s(m,N)   \right]^T \right),
\end{equation}
with $ p_s(i,j)$ being the probability of purification success at state $(i,j)$ known from~\eqref{eq:pur_success}.
Additionally, $\mathbf{D}_m$ includes the backward transitions due to a successful purification expressed in~\eqref{eq:backward_transitions}. 
We express the elements of the matrix representing the transition from state $(m,j)$ to state $(0,k)$ by
\begin{equation}
    \mathbf{D}_m[(m,j),(0,k)]=p_gp_s(m,j) \mathds{1}_{k=\triangle n_p(m,j)}, \ 0 \leq m \leq N ,
\end{equation}
where $\mathds{1}$ is the indicator function.

\subsection{Obtaining the fidelity distribution from the DTMC}

The classical steady-state solution to the DTMC to obtain the steady-state probability vector $\pmb{p}$ involves solving the linear system of equations $\pmb{p}^T \mathbf{Q}  = \pmb{p}^T$ with the normalization condition $\pmb{p}^T \pmb{e}_{n_s} = 1$, 
where $\mathbf{Q}$ is the transition matrix, $\pmb{e}_{n_s}$ is an all-one column vector of length ${n_s}$ while ${n_s}$ being the number of states.
Since the number of equations in the linear system grows quadratically as $O(N^2)$, we make use of the problem structure and derive next a reduced problem that requires solving only $N+1$ equations.
We denote the probability of a state $(i,j)$ as $p_{i,j}$ and the column probability vector of the block states $S_i$ as $\pmb{p}_i$.
Moreover, we denote the part of the transition matrix representing the transitions from all the states to the states $S_i$, i.e., a block column in $\mathbf{Q}$, by $\mathbf{Q}_i$.
For example $\mathbf{Q}_{-\infty}$ and $\mathbf{Q}_{0}$ represent the first and the second block column in $\mathbf{Q}$ as given in~\eqref{eq:transition_matrix}.
Using the steady-state description from above and \eqref{eq:transition_matrix}, we express $\pmb{p}_i$ in terms of $\mathbf{Q}_i$ as
\begin{equation}
    \pmb{p}^T \mathbf{Q}_i=\pmb{p}^T_i .
    \label{eq:steady_state_per_i}
\end{equation}

The key idea to reducing the system of equations to $N+1$ is by relating the steady-state probabilities of all the states in terms of $\pmb{p}_0$ using the structure of the DTMC and the transition matrix~\eqref{eq:transition_matrix}.
The structure of the DTMC implies that the states $S_0$ link all the states together.
First, the states $S_i, \ i >0$ recursively originate from the forward transitions of $S_0$ as given by the corresponding block columns $\mathbf{Q_i}$.
Equipped with this idea, we can recursively derive $\pmb{p}_i: i>0$ in terms of $\pmb{p}_0$ using \eqref{eq:transition_matrix} and \eqref{eq:steady_state_per_i}, i.e., the recursive structure starts from the third block column in \eqref{eq:transition_matrix}.
This recursive structure leads to
\begin{equation}
     \pmb{p}_{i}^T=\pmb{p}_{i-1}^T \mathbf{X}_{i-1}= \pmb{p}_{0}^T \prod_{m=0}^{i-1} \mathbf{X}_{m}, \ 0<i<N . 
     \label{eq:pr_i}
\end{equation}
Similarly, we derive ${p}_N$ as
 \begin{equation*}
     p_N=\pmb{p}^T_{N-1}  \mathbf{X_{N-1}} +(1-p_g)p_{N} =\frac{1}{p_g} \pmb{p}^T_{N-1} \mathbf{X_{N-1}} .
 \end{equation*}
 Note that $p_N$ represents only one state, i.e., $p_{N,N}$.
 We further derive ${p}_{N}$ using the expression of $\pmb{p}_{N-1}$ in terms of $\pmb{p}_{0}$ from \eqref{eq:pr_i} as
 \begin{equation}
    p_N=\frac{1}{p_g} \pmb{p}^T_{0} \prod_{m=0}^{N-1} {X_{m}} .
         \label{eq:pr_N}
 \end{equation}
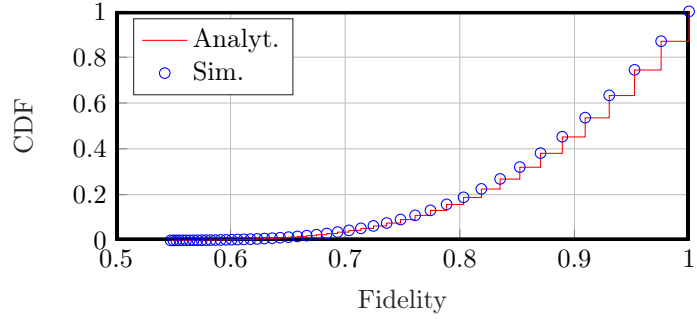
\begin{figure}[t]
    \centering
%
%
\begin{tikzpicture}

\begin{axis}[%
width=3in,
height=1.2in,
scale only axis,
xmin=0.5,
xmax=1,
xlabel style={font=\color{white!15!black}},
xlabel={Fidelity},
ymin=0,
ymax=1,
ylabel style={font=\color{white!15!black}},
ylabel={CDF},
axis background/.style={fill=white},
axis x line*=bottom,
axis y line*=left,
xmajorgrids,
ymajorgrids,
legend style={legend cell align=left, align=left, draw=white!15!black},
legend pos=north west
]
\draw[line width=0.005 \linewidth](current axis.south west)rectangle(current axis.north east);

\addplot[const plot, color=red] table[row sep=crcr] {%
0.547684581107775	0.000172152803854549\\
0.550129421861402	0.000209189254273189\\
0.552699612280932	0.000254158582749807\\
0.555401579181167	0.000308751370964108\\
0.558242078886748	0.000375016376088723\\
0.561228214126491	0.000455435895754235\\
0.564367451793902	0.000553016861664461\\
0.567667641618306	0.000671400912124751\\
0.571137035793257	0.000814997357254202\\
0.574784309611318	0.000989143747535133\\
0.578618583156814	0.00120029971288482\\
0.582649444110793	0.00145628088692739\\
0.586886971725223	0.00176654110687301\\
0.591341762026367	0.00214251272770674\\
0.596024954310377	0.00259801686290572\\
0.600948258997328	0.00314372077433569\\
0.606123986913372	0.00380267905992231\\
0.611565080074215	0.00459819451167662\\
0.617285144046899	0.00555826181322418\\
0.623298481970803	0.00671652537336405\\
0.629620130322946	0.00811368540094202\\
0.636265896517006	0.00979809738178604\\
0.643252398430095	0.0118276084301995\\
0.650597105956101	0.014271937481662\\
0.658318384689527	0.0172151572463451\\
0.66643554184904	0.0207570817494008\\
0.674968874555578	0.0250175025457812\\
0.683939720585721	0.0301398283177951\\
0.693370511727251	0.0362948975861308\\
0.7032848298703	0.0436872216021003\\
0.713707465974363	0.0525394474397516\\
0.724664482058611	0.0631481553380684\\
0.736183276370507	0.0758576252070359\\
0.748292651895705	0.0910704642883996\\
0.761022888380508	0.109266710504054\\
0.774405818047013	0.130997075435152\\
0.788474905190243	0.156933602224632\\
0.803265329856317	0.187739657037959\\
0.818814075810887	0.224388178578514\\
0.83516002301782	0.267888848846333\\
0.852344044859357	0.319493888064559\\
0.870409110340859	0.380223519311357\\
0.889400391535702	0.451774040927378\\
0.909365376538991	0.535205970133438\\
0.930353988212529	0.632573557849526\\
0.95241870901798	0.744045045255637\\
0.975614712250357	0.86925556080446\\
1	1\\
};
\addlegendentry{Analyt.}


\addplot[const plot, color=blue, draw=none, only marks, mark=o, mark options={solid, blue}] table[row sep=crcr] {%
0.547684581107775	0.0001635\\
0.550129421861402	0.0002025\\
0.552699612280932	0.000247\\
0.555401579181167	0.0003005\\
0.558242078886748	0.0003655\\
0.561228214126491	0.000449\\
0.564367451793902	0.000554\\
0.567667641618306	0.0006815\\
0.571137035793257	0.0008365\\
0.574784309611318	0.0010195\\
0.578618583156814	0.001232\\
0.582649444110793	0.0014925\\
0.586886971725223	0.0018005\\
0.591341762026367	0.002173\\
0.596024954310377	0.00263\\
0.600948258997328	0.0031865\\
0.606123986913372	0.0038585\\
0.611565080074215	0.0046755\\
0.617285144046899	0.0056585\\
0.623298481970803	0.006844\\
0.629620130322946	0.008265\\
0.636265896517006	0.009974\\
0.643252398430095	0.0120235\\
0.650597105956101	0.0144725\\
0.658318384689527	0.017422\\
0.66643554184904	0.020983\\
0.674968874555578	0.025256\\
0.683939720585721	0.030389\\
0.693370511727251	0.0365465\\
0.7032848298703	0.043923\\
0.713707465974363	0.0527815\\
0.724664482058611	0.063415\\
0.736183276370507	0.076161\\
0.748292651895705	0.0914255\\
0.761022888380508	0.1096805\\
0.774405818047013	0.1314175\\
0.788474905190243	0.1573945\\
0.803265329856317	0.1882505\\
0.818814075810887	0.2250175\\
0.83516002301782	0.2685925\\
0.852344044859357	0.320274\\
0.870409110340859	0.381117\\
0.889400391535702	0.4526585\\
0.909365376538991	0.536094\\
0.930353988212529	0.633326\\
0.95241870901798	0.7446345\\
0.975614712250357	0.869689\\
1	1\\
};
\addlegendentry{Sim.}

\end{axis}

\end{tikzpicture}%
    \caption{The simulation of a link with $l=15 \si{\km}$ validates the analytical distribution of the fidelity of the older EPR pair as calculated from the DTMC model. }
    \label{fig:CMF_validation}
\end{figure}

Now, the state $(-\infty,0)$ is the destination of the states $S_i, \ i\geq 0$ as a result of the backward transitions capturing the failed purification attempt which is represented by the first column in $\mathbf{Q}$. 
Therefore, using \eqref{eq:steady_state_per_i}, we derive ${p}_{-\infty,0}$ in terms of $\pmb{p}_{i}, \ i\geq 0$ as
\begin{equation*}
    p_{-\infty,0}= \sum_{m=0}^N \pmb{p}_m^T \pmb{f}_m = \pmb{p}_0^T \pmb{f}_0+\pmb{p}_N^T \pmb{f}_N +\sum_{m=1}^{N-1} \pmb{p}_m^T \pmb{f}_m.
\end{equation*}
Consequently, using the expressions in \eqref{eq:pr_i} and \eqref{eq:pr_N} we obtain
\begin{align}
    p_{-\infty,0} &=\pmb{p}_0^T \left[ \pmb{f}_0+ \frac{1}{p_g} \prod_{m=0}^{N-1} \mathbf{X}_m \pmb{f}_N + \sum_{m=1}^{N-1} \prod_{n=0}^{m-1} \mathbf{X}_n \pmb{f}_m \right] 
       , \nonumber \\
    &:= \pmb{p}_0^T \mathbf{\Phi}.
 \end{align}
Next, the states $S_{-\infty}$ are recursively related by the forward transitions according to $\mathbf{Q}_{-\infty}$ as  
 \begin{align*}
    p_{-\infty,j}&= (1-p_g) p_{-\infty,j-1}=(1-p_g)^j p_{-\infty,0} , \ 0<j<N 
    , \nonumber \\
     p_{-\infty,N}&=\frac{1-p_g}{p_g} p_{-\infty,N-1}
    =\frac{(1-p_g)^N}{p_g} p_{-\infty,0} .
 \end{align*}
Let $\pmb{\rho}=\left[ 1,(1-p_g), \dots,{(1-p_g)}^{N-1},{{(1-p_g)}^{N}}/{p_g} \right]^T$,
we rewrite $ \pmb{p}_{-\infty}$ in vector form in terms of $\pmb{p}_{0}$ as
 \begin{equation}
     \pmb{p}_{-\infty}^T= p_{-\infty,0}\ \pmb{\rho}^T= \pmb{p}_{0}^T \mathbf{\Phi} \pmb{\rho}^T .
     \label{eq:pr_inf}
 \end{equation}

Finally, $S_0$ is the destination of all the states according to $\mathbf{Q}_0$, i.e., from $S_{-\infty}$ according to the forward transitions in \eqref{eq:forward_transitions} and from all the other states according to the backward transitions due to successful purification in \eqref{eq:backward_transitions}. Therefore, we describe this relation using $\eqref{eq:steady_state_per_i}$ as
  \begin{equation}
     \pmb{p}_0^T=\pmb{p}_{-\infty}^T \mathbf{X}_{-\infty}+ \sum_{m=0}^{N} \pmb{p}_{m}^T \mathbf{D}_{m} .
     \label{eq:pr_0}
 \end{equation}
As a result, the linear system of equation to be solved is reduced to 
\begin{align}
    \pmb{p}_0^T \mathbf{\Psi}&= \pmb{0}_{N+1,1} , \nonumber \\
        \pmb{p}_0^T \pmb{\beta}&= 1 ,
        \label{eq:reduced_eqns}
\end{align}
where we derive $\mathbf{\Psi}$ using \eqref{eq:pr_i}, \eqref{eq:pr_N} and \eqref{eq:pr_inf} in  \eqref{eq:pr_0} as
 \begin{align}
     \mathbf{\Psi}:=  
     \mathbf{I}_N-\mathbf{\Phi} \pmb{\rho}^T \mathbf{X}_{-\infty}-\mathbf{D}_{0}-\sum_{m=1}^{N-1} \prod_{n=0}^{m-1} \mathbf{X}_n \mathbf{D}_m
     -\frac{1}{p_g} \prod_{m=0}^{N-1} \mathbf{X}_m \pmb{D}_N ,
 \end{align}
 in addition to $\pmb{\beta}$ using \eqref{eq:pr_i}, \eqref{eq:pr_N} and \eqref{eq:pr_inf} in the normalization equation as
 \begin{equation}
     \pmb{\beta}:= \mathbf{\Phi} \pmb{\rho}^T \pmb{e}_{N+1}+\pmb{e}_{N+1}+\sum_{m=1}^{N-1} \prod_{n=0}^{m-1} \mathbf{X}_n \pmb{e}_{N-m+1}+\frac{1}{p_g} \prod_{m=0}^{N-1} \mathbf{X}_m .
 \end{equation}

We rewrite \eqref{eq:reduced_eqns} in a short form as
 \begin{equation}
    \pmb{p}_0^T \left[\mathbf{\Psi} | \pmb{\beta} \right]= \left[\pmb{0}_{1,N+1}|1\right] .
    \label{eq:reduced_eqns_compact}
 \end{equation}
using  the column-wise concatenation operation  $[.|.]$.

The linear system of equations in \eqref{eq:reduced_eqns_compact} is of rank $N+1$, where its solution yields the value of $\pmb{p}_0$.
In addition, we obtain the other steady-state probabilities by substituting $\pmb{p}_0$ in \eqref{eq:pr_i}, \eqref{eq:pr_N} and \eqref{eq:pr_inf}.
\section{Numerical Validation}
\label{sect:numerical_validation}

In this section, we validate our DTMC analytical approach with simulations and show the trade-off between the steady-state average fidelity of the stored EPR pairs defined as $\bar{F_i}:=\underset{n \rightarrow \infty}{\lim} \E[F_i(n)]$ and their average number for an increasing link length ranging between $5 \si{\km}$ and $30 \si{\km}$. 
We set the attenuation $\eta=0.15\ \si{\dB / \km}$ and the decoherence time $t_c=1 \ \si{\ms}$ similar to~\cite{Towsley_opt_ent_pur, Netsquid}. 
We assume a perfect generation of EPR pairs and use a fidelity threshold $F_\epsilon=0.55$. 

In Fig.~\ref{fig:CMF_validation}, we validate the steady-state analytical cumulative mass function (CMF) of the older EPR pair with the result from the simulation for $l=15 \si{\km}$. 

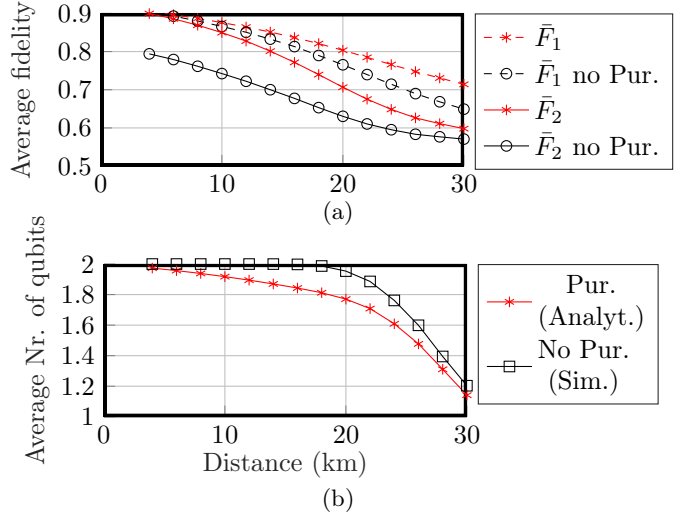
\begin{figure}[t]
\centering
      \begin{subfigure}[b]{1\linewidth}
            \centering
        {
%
%
\begin{tikzpicture}

\begin{axis}[%
width=1.9 in,
height=0.8 in,
scale only axis,
xmin=0,
xmax=30,
xlabel style={font=\color{white!15!black}},
xlabel={},
ymin=0.5,
ymax=0.9,
ylabel style={font=\color{white!15!black}},
ylabel={Average fidelity},
 xlabel style={yshift=0.2cm},
 ylabel style={yshift=-0.2cm},
axis background/.style={fill=white},
axis x line*=bottom,
axis y line*=left,
xmajorgrids,
ymajorgrids,
legend style={legend cell align=left, align=center, draw=white!15!black},
legend pos=outer north east
]
\draw[line width=0.005 \linewidth](current axis.south west)rectangle(current axis.north east);
\addplot [color=red, dashed, mark=asterisk, mark options={solid, red}]
  table[row sep=crcr]{%
4	0.90549388273468\\
6	0.896551855804589\\
8	0.886754126839962\\
10	0.876006299783752\\
12	0.864170899928125\\
14	0.851185145183572\\
16	0.836796218412207\\
18	0.820873157095249\\
20	0.803461240403479\\
22	0.784786286481158\\
24	0.766038997966124\\
26	0.747500174915002\\
28	0.730825884489367\\
30	0.714204178236461\\
};
\addlegendentry{$\bar{F}_1$ }


\addplot [color=black,dashed, mark=o, mark options={solid, black}]
  table[row sep=crcr]{%
4	0.903069518044919\\
6	0.892845242144034\\
8	0.880023103068945\\
10	0.865821550814653\\
12	0.850551086275231\\
14	0.832594522676336\\
16	0.813049444722067\\
18	0.789743094239421\\
20	0.765406084769243\\
22	0.739740574332198\\
24	0.71446670361614\\
26	0.689875861560955\\
28	0.668466117826107\\
30	0.649179226935605\\
};
\addlegendentry{$\bar{F}_1$ no Pur.}

\addplot [color=red, mark=asterisk, mark options={solid, red}]
  table[row sep=crcr]{%
4	0.899525052871245\\
6	0.885762444734514\\
8	0.869083512368917\\
10	0.849605741805453\\
12	0.827520204497579\\
14	0.801080014185385\\
16	0.771717369812174\\
18	0.739750531629751\\
20	0.706599058454884\\
22	0.675343198126924\\
24	0.64852699054507\\
26	0.626304329279537\\
28	0.611232470697052\\
30	0.598112802273894\\
};
\addlegendentry{$\bar{F}_2$}


\addplot [color=black, mark=o, mark options={solid, black}]
  table[row sep=crcr]{%
4	0.794015077025132\\
6	0.779149016462997\\
8	0.761491021529991\\
10	0.742365374233937\\
12	0.722260173510602\\
14	0.700093198082794\\
16	0.677348398040609\\
18	0.653160661930324\\
20	0.630394699208167\\
22	0.610431862126463\\
24	0.595272018097537\\
26	0.583355229478481\\
28	0.576708252600698\\
30	0.570755424037249\\
};
\addlegendentry{$\bar{F}_2$ no Pur.}

\end{axis}
\end{tikzpicture}%
}
        \captionsetup{skip=-1pt} 
        \caption{}
        \label{fig:average_fidelity}
     \end{subfigure}
         \\
     \vspace{-10pt}
    \begin{subfigure}[b]{1\linewidth}
            \centering
                 \hspace{10cm}
         {
%
%
\begin{tikzpicture}

\begin{axis}[%
width=1.9in,
height=0.8in,
scale only axis,
xmin=0,
xmax=30,
xlabel style={font=\color{white!15!black}},
xlabel={Distance (\si{\km})},
ymin=1,
ymax=2,
ylabel style={font=\color{white!15!black}},
ylabel={Average Nr. of qubits},
axis background/.style={fill=white},
axis x line*=bottom,
axis y line*=left,
xmajorgrids,
ymajorgrids,
 xlabel style={yshift=0.2cm},
 ylabel style={yshift=-0.35cm},
legend style={legend cell align=left, align=center, draw=white!15!black},
legend pos= outer north east
]
\draw[line width=0.005 \linewidth](current axis.south west)rectangle(current axis.north east);
\addplot [color=red, mark=asterisk, mark options={solid, red}]
  table[row sep=crcr]{%
4	1.97335772143871\\
6	1.95700662584734\\
8	1.93835659501354\\
10	1.9174789431222\\
12	1.89470783262379\\
14	1.86965529189218\\
16	1.84286186898684\\
18	1.81205773329749\\
20	1.77037175489266\\
22	1.70835700449618\\
24	1.60865266804506\\
26	1.47716281957811\\
28	1.31076377671788\\
30	1.14133765499133\\
};
\addlegendentry{Pur. \\(Analyt.)}


\addplot [color=black, mark=square, mark options={solid, black}]
  table[row sep=crcr]{%
4	1.999997\\
6	1.999999\\
8	1.999998\\
10	1.9999995\\
12	1.999973\\
14	1.999737\\
16	1.9977025\\
18	1.987176\\
20	1.9537725\\
22	1.8860175\\
24	1.7617885\\
26	1.598908\\
28	1.3947565\\
30	1.203239\\
};
\addlegendentry{No Pur. \\ (Sim.)}

\end{axis}
\end{tikzpicture}%
}
          \captionsetup{skip=0pt} 
        \caption{}
        \label{fig:average_qubits}
     \end{subfigure} 
          \captionsetup{skip=0pt} 
\caption{The trade-off between the average number of stored EPR pairs and their fidelity by applying purification on the stored qubits for an increasing link distance: (a) The analytical average fidelity of the two EPR pairs $\bar{F_1} $ and $\bar{F_2} $ in case of applying purification is larger. (b) The average number of the stored EPR pairs is, however, smaller in the case of applying purification.}
    \label{fig:rate_fidelity_tradeoff}
    \vspace{-15pt}
\end{figure}

We illustrate in Fig.~\ref{fig:rate_fidelity_tradeoff} the rate-fidelity trade-off achieved by applying \textit{purification beyond generation} to the stored EPR pairs in our system with two quantum memories.
Intuitively, while purification improves the average steady-state fidelity of the two stored EPR pairs as shown in Fig.~\ref{fig:average_fidelity}, it results in a reduction in the average number of the EPR pairs as shown in Fig.~\ref{fig:average_qubits} since we sacrifice one EPR pair for successful purification and both in case of failure.
Note that $\bar{F}_1$ represents the average fidelity of the higher fidelity EPR pair when it exists, i.e., when the quantum memories are full.



\section{Related Work}
\label{sect:related_work}
Link-level entanglement is the first step towards long distant quantum communication. 
The authors of~\cite{dahlberg_qu_Linllayer_protocol} propose a physical and link layer protocol to provide a robust link-level entanglement generation between quantum communication nodes.
Specifically, the proposed protocol organizes the link-level entanglement generation requests to ensure the fidelity desired by the applications at the expense of the increased generation time.
Nitrogen vacancy (NV) centers in diamond platform~\cite{Riedel_Nitrogen_vacancy} is one way to generate desired fidelity EPR pairs, where higher fidelity EPR pairs require longer generation times. 
A different method relies on recurrence purification algorithms, which use two EPR pairs per round to obtain a higher fidelity one.
The work in~\cite{Ruan_pur_protocol} proposes an approach that purifies two EPR pairs using polarization mode dispersion and derives an expression of the improved fidelity as well as the probability of purification success.
Several other works such as~\cite{Inside_quantum_repeater,deutsch_quantum_pur} provide quantum operation-based procedures for the purification of two EPR pairs.

Starting from the Lindblad formalization of the qubit interaction with the environment, i.e., decoherence, as time first order differential equation~\cite{schlosshauer_decoherence}, the time dynamics of the fidelity can be analytically expressed for different phase damping models~\cite{cacciapuoti_qu_decoherence}.
Using this concept, the works in~\cite{Inside_quantum_repeater} express the exponentially decaying fidelity over time of the EPR pairs.
Hence, quantum communication nodes need to address the effect of the decoherence on the stored link-level EPR pairs by estimating their fidelity to ensure meeting the desired application requirements.
For that reason, the works in~\cite{Netsquid,Towsley_stochastic_qu_swithc } drop qubits from the memory after specific cut-off times to ensure a minimum fidelity requirement.
Specifically, the authors in~\cite{Towsley_stochastic_qu_swithc} probabilistically model the cut-off times by an exponential distribution.
On the other hand, the work in~\cite{Dai_Qu_Queuing_delay} models a quantum queue without dropping qubits and derives an expression on the average queuing delay, thus it can estimate the average decoherence a qubit suffers in the queue.
Overall, these works differ from this paper in the sense that we target the derivation of the \textit{steady-state distribution} of the fidelity of EPR pairs on one link given a continuous purification after generation protocol.




\section{Discussion \& Open problems}
\label{sect:Discussion}
In this paper, we used a DTMC to model the fidelity of the EPR pairs for a quantum communication link in a few (two) quantum memory system.
We used this model to calculate the steady-state distribution of the fidelity of the EPR pairs. 
The model shows the improvement of the fidelity in terms of its distribution of the existing EPR pairs by applying a purification beyond generation protocol at the expense of a decrease in the average number of ready EPR pairs in the system.

Extending the model to more than two quantum memories or a quantum memory queue is open for future work as well as incorporating a request process that consumes the EPR pairs as required by the desired application.
Moreover, having more than a few EPR pairs stored in the queue raises a question about the appropriate purification beyond generation protocol and when it should be applied.
Further, the problem of calculating the distribution of the continuous fidelity is open and is considered much more complex due to the stochastic behavior of the entanglement generation and purification as well as the dependence between the fidelity at the purification points resulting in random recursive equations.

\balance
\bibliographystyle{IEEEtran.bst}
\bibliography{IEEEabrv,bibliocache.bib}
\end{document}